\def\dev{\partial}
\def\G{\Gamma}
\def\m{\mu}
\def\n{\nu}

\def\l{\lambda}
\def\T{\Theta}

\def\pmb#1{\setbox0=\hbox{#1}%
  \kern-.025em\copy0\kern-\wd0
  \kern.05em\copy0\kern-\wd0
  \kern-.025em\raise.0433em\box0}

\def\bfxi{\pmb{$\xi$}}
\documentstyle[aps,preprint]{revtex}
\topmargin = - 0.9 cm
\begin{document}
\preprint{IFUP-TH-60/92\newline\ \
 December 1992}
\title{
 Closed time like curves and the energy condition in 2+1
dimensional gravity
}
\author{P. Menotti}
\address{
Dipartimento di Fisica della Universit\`a, Pisa 56100, Italy and\\
INFN, Sezione di Pisa
}
\author{D. Seminara}
\address{
Scuola Normale Superiore, Pisa 56100, Italy and\\
INFN, Sezione di Pisa}
\maketitle
\begin{abstract}
We consider gravity in 2+1 dimensions in presence of extended
stationary sources with rotational  symmetry. We prove by direct use
of Einstein's equations that if  i) the energy momentum tensor
satisfies the weak energy condition, ii) the universe is open
(conical at space infinity), iii) there are no CTC at space infinity,
then there are no CTC at all.
\end{abstract}
The problem of closed time like curves (CTC) in 2+1 dimensional gravity has
recently attracted considerable interest
\cite{Gott,JDtH,GuthI,GuthII,Cutler,Kabat,Ori,'t Hooft,Tipler,Hawking,DJ}.
Gravity in 2+1 dimensions in
addition
to describing solutions of 3+1 dimensional gravity  in presence of cosmic
strings, is of interest in itself because  due to its
simplicity
can be analyzed in great detail.
Gott \cite{Gott}
 was able to produce examples of kinematical configurations of point
particles without spin which produce CTC in 2+1 dimensions.
This system came under close
scrutiny
\cite{JDtH,GuthI,GuthII,Cutler,Kabat,Ori,'t Hooft}.
 Carrol, Fahri and Guth \cite{GuthII} prove that if the universe
containing  spinless point particles is open and has
total time like momentum, no Gott pair can be created during its evolution
thus supporting the idea that in an open timelike universe no CTC can form.
't Hooft \cite{'t Hooft} for a system of spinless point particles, using
a general construction of a complete series of time ordered Cauchy
surfaces, concludes that, also in the case of a closed universe, if
Gott pairs are produced
as envisaged in \cite{GuthII}, the universe collapses before a CTC
can form. Tipler \cite{Tipler} and Hawking \cite{Hawking} assuming the
weak energy condition proved in
3+1 dimensions that if CTC are formed in a compact region of space-time,
then one must necessarily have the creation of  singularities.
On the other hand it is very simple to recognize \cite{JDtH}
that a point like source
with angular momentum produce sufficiently near the source CTC. The existence
of CTC appears to generate some irresolvable clash with causality
\cite{Hawking}. In the
case of  the point like spinning source the appearance of CTC is usually
ascribed to  the unphysical nature of a point like spinning object \cite{DJ}.
 A conjecture \cite{JDtH,Hawking} states that if the metric is
generated by a physical energy momentum tensor and proper boundary
conditions are imposed, no CTC should develop. By physical sources one usually
understands an energy momentum tensor which satisfies one or more among the
weak , dominant or strong energy condition \cite{H.E.}.
For this reason  we  investigate the problem in 2+1 dimensions
in presence of extended
sources in the simple instance of a stationary source with rotational
symmetry. Our result  is the following: if the energy momentum tensor
satisfies the weak energy condition (WEC),
the universe is open (conical at
space infinity), and there are not CTC at space infinity, then there are
no CTC at all. The proof follows from direct manipulation of Einstein' s
equation combined with the WEC. The proof immediately extends to the case of
cylindrical universes with angular momentum and to the case of ``linear''
universes.\\
In the treatment we shall extensively use the reduced radial gauge defined
in  \cite{MS1} and the method of solution developed in \cite{MS2}.
In stationary  problems the reduced radial gauge \cite{MS1} is defined by
\begin{equation}
\sum_i\xi^i\G^a_{bi}=0\ \ , \ \ \sum_i\xi^i e^a_i=\sum_i \xi^i\delta^a_i
\end{equation}
where the sums run over the space indices. (In the following the indices $i
,\ j,\ k,\ l,\ m$ run over space indices).
In any space time dimensions there are resolvent formulae \cite{MS1}
 which express the
vierbeins and the connection  in terms of the Riemann tensor in this
gauge; they are analogous to those for the complete radial gauge
\cite{MT}
\begin{equation}
\Gamma^a_{bi}({\bfxi})=\xi^j \int^1_0 R^a_{bji}(\l{\bfxi})\l d\l\ \ ,\ \
\Gamma^a_{b0}({\bfxi})=\G^a_{b0}({\bf 0})+\xi^i\int^1_0 R^a_{bi0}(\l {\bfxi})
d\l
\end{equation}
\begin{equation}
\label{eam}
e^a_i=\delta^a_i+\xi^l\xi^j\int^1_0 R^a_{jli}(\l{\bfxi})\l(1-\l) d\l\ \ ,\ \
e^a_0=\delta^a_0+\xi^i\G^a_{i0}({\bf 0})+\xi^i\xi^j\int^1_0 R^a_{ij0}
(\l{\bfxi})(1-\l) d\l
\end{equation}
The simplifying feature
 of 2+1 dimensions is the substantial identification of the
Riemann with the Ricci tensor and through Einstein's equation, with the
energy momentum tensor. Explicitly, denoting with $R^{ab}$ the Riemann
two-form we have
\begin{equation}
\label{equazione}
R^{ab}=-{\kappa\over 4}\varepsilon^{abc} T_c=-{\kappa\over 8}
\varepsilon^{abc}~\varepsilon_{\rho\m\n}\tau^{~\rho}_c dx^\m\wedge dx^\n,
\end{equation}
where $T_c$ is the energy momentum two form. Using such a relation one can
express through  a simple quadrature, the connections and  the
dreibeins in terms of the energy momentum tensor which is the source of
the gravitational  field and thus one solves Einstein's equation.
On the other hand the energy momentum tensor
  is  subjected to the covariant conservation
law and  symmetry condition. Thus our problem is to construct
the general form of
conserved symmetric sources in the reduced radial gauge  which in addition
should satisfy other physical requirements given by the support of the
sources and the restrictions given by the energy condition \cite{H.E.}.\\
The conservation and symmetry equations are
\begin{equation}
{\cal D} T^a=0\ \ ,\ \ \varepsilon_{abc} T^b\wedge e^c =0.
\end{equation}
To solve these equations  we use the technique  developed in \cite{MS2}.
We introduce the cylindrical coordinates
\begin{equation}
\label{coordinates}
\xi^0=t\ \ ,\ \ \xi^1=\rho\cos\theta\ \ ,\ \ \xi^2=\rho\sin\theta
\end{equation}
and the cotangent vectors
\begin{equation}
\label{vectors}
T_\mu=\frac{\partial \xi^0}{\partial \xi^\mu}\ \ ,\ \
\Theta_\mu=\rho\frac{\partial\theta}{\partial\xi^\mu}\ \ {\rm and}\ \
P_\mu=\frac{\partial\rho}{\partial \xi^\mu}.
\end{equation}
In 2+1 dimensions the most general form of a radial $\Gamma^{ab}_\mu(\bfxi)$
is obviously given by
\begin{equation}
\Gamma^{ab}_\mu(\bfxi)=\varepsilon^{abc}\varepsilon_{\mu\rho\nu}P^\rho
A^\nu_c(\bfxi)
\end{equation}
and we parameterize $A^\nu_c(\bfxi)$ as follows
\begin{equation}
\label{A}
A^\rho_c(\bfxi)= T_c \left [\beta_1 T^\rho+\frac{(\beta_2-1)}{\rho}\Theta^\rho
\right ]+\Theta_c \left [\alpha_1 T^\rho+\frac{\alpha_2}{\rho}\Theta^\rho
\right ]+P_c \left [\gamma_1 T^\rho+\frac{\gamma_2}{\rho}\Theta^\rho
\right ].
\end{equation}
We use for the coefficient of $T_c \Theta^\rho$ the form $\beta_2-1$ as it
simplifies the writing of subsequent formulae.
Using (\ref{A}) and (\ref{eam}) we obtain for the dreibeins
\begin{equation}
\label{drei1}
e^a_0(\bfxi)=-T^a A_1 -\Theta^a  B_1\ \ , \ \
e^a_i(\bfxi)=-\frac{1}{\rho}\Theta^a\Theta_i B_2-
\frac{1}{\rho} T^a \Theta_i A_2- P^a P_i
\end{equation}
where $A_i$ and $B_i$ are defined by
\begin{equation}
A_1(\bfxi)=\rho \int^1_0 \alpha_1(\lambda \bfxi) d\lambda-1\ \  , \  \
A_2(\bfxi)=\rho \int^1_0 \alpha_2(\lambda \bfxi) d\lambda\ \  ,  \
B_i(\bfxi)=\rho \int^1_0 \beta_i(\lambda \bfxi) d\lambda
\end{equation}
from which the metric is given by
\begin{equation}
ds^2= (A_1^2-B_1^2) dt^2+2 (A_1 A_2-B_1 B_2)d t d\theta+
(A_2^2-B^2_2)d\theta^2 -d r^2.
\end{equation}
Substituting (\ref{A}) in Einstein's equation (\ref{equazione})
one obtains the most general energy momentum tensor in the
radial gauge\cite{MS1,MS2}.
\begin{eqnarray}
\label{tau1}
&&\tau^{a\rho}=-\frac{4}{\kappa}\biggl \{ T^a \biggl ( T^\rho
\frac{\beta^\prime_2}{\rho}+\T^\rho \beta^\prime_1\biggr )+
\T^a
\biggl ( T^\rho
\frac{\alpha^\prime_2}{\rho}+\T^\rho\alpha^\prime_1
\biggr )+
 P^a\biggl (T^\rho \frac{\gamma^\prime_2}{\rho}+
\T^\rho
\gamma^\prime_1\biggr ) +\\
&&\frac{1}{\rho}P^\rho
\biggl [ T^a\biggl (\alpha_1\gamma_2-\alpha_2\gamma_1-
\frac{\dev\beta_1}{\dev \theta}\biggr )+
\T^a \biggl (\beta_1\gamma_2-\beta_2\gamma_1-
\frac{\dev\alpha_1}{\dev \theta}\biggr )
+P^a \biggl (\alpha_1\beta_2
-\alpha_2\beta_1-
\frac{\dev\gamma_1}{\dev \theta}\biggr )\biggr ]\biggr \}.\nonumber
\end{eqnarray}
where the primes denote derivatives with respect to $\rho$. The vanishing
of $\tau^\rho_c$ outside the source i.e. for $\rho\ge\rho_0$ implies that in
such a region $\alpha_i,\ \beta_i,\ \gamma_i$ are constant in $\rho$ and
the coefficients of $P^\rho T_c,\ P^\rho \Theta_c$ and $P^\rho P_c$ vanish.
In the case of rotational symmetry i.e. when
$\alpha_i,\ \beta_i,\ \gamma_i$ do not depend on $\theta$
\cite{MS2}, it is easily shown that the symmetry condition
on $\tau$ imposes $\gamma_i=0$, and the contraction of $\tau^{a\mu}$ on the
dreibein gives
\begin{eqnarray}
\label{tab}
&&\tau^{a\mu}(\bfxi) e^c_\mu(\bfxi)=\tau^{ac}=-\frac{4}{\kappa \rho}\biggl \{
T^a \biggl [ T^c \biggl (A_2 \beta^\prime_1-A_1 \beta^\prime_2 \biggr )+
 \Theta^c \biggl ( A_2 \alpha_1^\prime-A_1\alpha_2^\prime\biggr )\biggr ]+
\nonumber \\
&&\Theta^a \biggl [ T^c \biggl (B_2 \beta^\prime_1-B_1 \beta^\prime_2
\biggr )+
\Theta^c \biggl ( B_2 \alpha_1^\prime-B_1\alpha_2^\prime\biggr )\biggr ]+
P^a P^c\biggl (\alpha_1 \beta_2-\alpha_2 \beta_1\biggr )\biggr \}.
\end{eqnarray}
The only surviving symmetry condition becomes
\begin{equation}
\label{seq}
A_1\alpha_2^\prime -A_2 \alpha_1^\prime +B_2 \beta_1^\prime -B_1
\beta_2^\prime=0.
\end{equation}
Eq. (\ref{seq}) is easily integrated; the regularity condition
on $\Gamma^{ab}_\mu (\bfxi)$ at the origin imposes $\alpha_1=o(\rho),\
\alpha_2=o(\rho),\ \beta_1=O(1) \ {\rm and}\  \beta_2=1+o(\rho)$,
which implies that the integration constant is 0 to get
\begin{equation}
\label{symmetry}
A_1\alpha_2-A_2 \alpha_1+B_2\beta_1-B_1
\beta_2=0.
\end{equation}
It is immediately  seen that a CTC implies the existence of a CTC with
constant $\rho$ and $t$. In fact if we call $\lambda$ the parameter of the
curve $(0\le\lambda<1)$,  $t(\lambda)$ must satisfy $t(0)=t(1)$ and at the
point
$\lambda_0$ where $\displaystyle{{dt\over d\lambda}=0}$
we have that the tangent vector
$(0,d\theta,d\rho)$ is time like i.e. also $(0,d\theta,0)$ is time like,
and thus the
circle $t=0$ , $\rho=\rho(\lambda_0)$ is timelike.
Thus to prove that CTC cannot exist
is sufficient to prove that $g_{\theta\theta}$ which by
assumption is negative at space
infinity, cannot change sign. If the determinant of
the dreibein in the radial gauge, where
$g_{\rho\rho}\equiv
-1$, vanishes for a certain ${\bar{\rho}}$, it means that the manifold defined
by $\rho=\rm{const.}$ becomes one dimensional at $\rho=\bar\rho$, implying
that at $\bar\rho$ the universe closes.
Thus for an open universe we have that $\det(e)>0$ for
$\rho>0$;
in particular if we denote with $\rho_0$ a point outside the source we must
have $\displaystyle{
 \frac{d {\rm det}(e)}{d\rho}\mid_{\rho=\rho_0}\ge 0}$
because  for $\rho\ge \rho_0$, due to the support conditions, the determinant
is a linear  function of $\rho$ for $\rho\ge\rho_0$.
  We show now that the WEC combined with $\det(e)>0$ and the absence
of CTC at infinity imply the absence of CTC for any $\rho$. In fact
the WEC  $v_a \tau^{ab} v_b\ge 0$ applied to the vectors $(1,1,0)$ and
$(1,-1,0)$ gives
\begin{equation}
\frac{d}{d\rho} \left [ (\alpha_2\pm\beta_2)(B_1\pm A_1)-
                 (B_2\pm A_2)(\alpha_1\pm\beta_1)\right ]\ge 0
\end{equation}
which can be integrated between $\rho$ and any point $\rho_0$ outside
the source to give
\begin{eqnarray}
\label{WECdelta}
E^{(\pm)} (\rho)\equiv
(B_2\pm A_2)(\alpha_1\pm\beta_1)-
&&(\alpha_2\pm\beta_2)(B_1\pm A_1)\ge\nonumber\\
&&(B^0_2\pm A^0_2)(\alpha^0_1\pm\beta^0_1)-
(\alpha^0_2\pm\beta^0_2)(B^0_1\pm A^0_1)
\end{eqnarray}
We study now the sign of the r.h.s.. We have the following cases:\\
\medskip
\noindent i) \ $(\alpha^0_2)^2-(\beta^0_2)^2\not = 0$  (conical universe).\\
Under the hypothesis that there are no CTC at infinity, as outside the
source $g_{\theta\theta}$ becomes a quadratic polynomial in $\rho$, we have
\begin{equation}
\label{B}
(\alpha^0_2)^2-(\beta^0_2)^2<0.
\end{equation}
The support equation for $\tau^{\rho\rho}$ is
\begin{equation}
(\alpha^0_2-\beta^0_2)(\alpha^0_1+\beta_1^0)-
(\alpha^0_2+\beta^0_2)(\alpha^0_1-\beta_1^0)=0
\end{equation}
and using the equation of motion written in the form
\begin{eqnarray}
\label{E}
(A_2+B_2)(\alpha_1-\beta_1)+&&(A_2-B_2)(\alpha_1+\beta_1)
-\nonumber\\
&&(\alpha_2-\beta_2)(A_1+B_1)-(\alpha_2+\beta_2)(A_1-B_1)=0
\end{eqnarray}
and the determinant written in the form
\begin{equation}
{\rm det}(e)=\frac{1}{2} [ (A_2-B_2)(A_1+B_1)-(A_2+B_2)(A_1-B_1)],
\end{equation}
the r.h.s. of (\ref{WECdelta}) becomes
\begin{equation}
E^{(\pm)}(\rho_0)=-\frac{\alpha^0_2\pm\beta^0_2}{\alpha^0_2\mp\beta_2^0}
\frac{d}{d\rho} {\rm det}(e)\mid_{\rho=\rho_0}
\end{equation}
which due to (\ref{B}) and
$\displaystyle{\frac{d {\rm det}(e)}{d\rho}\mid_{\rho=\rho_0}\ge 0}$
is non negative i.e. $E^{(\pm)}(\rho_0)\ge 0$.\\
\medskip
\noindent ii) $\alpha_2=\beta_2\not=0$ (linear universe).\\
Using the support equation we have $\alpha_1^0=\beta_1^0$ which implies
$E^{(-)}(\rho_0)=0$. Using the equation of motion (\ref{E}) and
${\rm det}(e)\not =0$ one has $A^0_2-B^0_2\not =0$ and thus we obtain
\begin{equation}
E^{(+)}(\rho_0)=-2\frac{\alpha_2^0+\beta^0_2}{A^0_2-B^0_2} {\rm det }(e).
\end{equation}
But ${\rm det}(e)>0$ while the sign of $\displaystyle{
\frac{\alpha_2^0+\beta^0_2}{A^0_2-B^0_2}}
$ has to be negative if there are no CTC at infinity, because $A^2_2-B^2_2$
in this case behaves linearly at infinity.\\
Similarly for $\alpha_2=-\beta_2$ we obtain $E^{(+)}(\rho_0)=0$
and $E^{(-)}(\rho_0)>0$.\\
\medskip
\noindent iii) If $\alpha^0_1=\alpha_2^0=\beta^0_1=\beta^0_2=0$
(cylindrical universe with angular momentum)
we have also $E^{(\pm)}(\rho_0)=0$.
\\
The only case which escapes our analysis is
$\alpha_2^0=\beta^0_2=0$ and $\alpha^0_1$ and/or $\beta^0_1$ $\not = 0$;
this situation corresponds to a cylindrical universe generated by a string
with tension and total angular momentum 0. Thus except for this case  we
have that the r.h.s. of equation (\ref{WECdelta}) is
$E^{(\pm)}(\rho_0)\ge 0$. Let us now consider the following combination
\begin{equation}
(A_2-B_2)^2 E^{(+)}(\rho)+(A_2+B_2)^2 E^{(-)}(\rho)\ge 0.
\end{equation}
A little algebra shows that the l.h.s. equals
\begin{equation}
-2 {\rm det}^2(e) \frac{d}{d\rho} \left (\frac{ A^2_2(\rho)-
B^2_2(\rho)}{{\rm det}(e)}\right ).
\end{equation}
Thus we reached the conclusion that $\displaystyle{\frac{d}{d\rho}
\left (\frac{ A^2_2(\rho)-
B^2_2(\rho)}{{\rm det}(e)}\right )}\le 0$; which means  that
$\displaystyle{
\left (\frac{ A^2_2(\rho)-
B^2_2(\rho)}{{\rm det}(e)}\right )}$ is a non increasing function
of $\rho$. As $\displaystyle{
\left (\frac{ A^2_2(\rho)-
B^2_2(\rho)}{{\rm det}(e)}\right )}$  at the origin is zero we obtain
that $A^2_2(\rho)-
B^2_2(\rho)$ is always negative and thus we cannot have CTC.\\
It is not difficult to produce examples in which all energy conditions are
satisfied
and CTC exist for any radius larger than a given one; but we obviously violate
the
condition that no CTC exists at infinity. Also renouncing to the energy
conditions
one can produce examples \cite{MS1} in which non CTC exist at infinity, but
they
exist at
finite radius. As mentioned above
the only case of open universe  with no CTC at infinity
which
escapes our analysis is the rather unphysical situation of a cylindrical
universe with
zero angular momentum of the type generated by a closed string with tension
\cite{classsol}
for which on the other hand no CTC exists outside the source. We have no
example of
an extended source satisfying the energy condition and producing such a
universe.\\
We mention that the approach with the radial gauge,
for which the solution can be given in terms
of
quadratures\cite{MS1,MS2},
 works also for the case of
rotational symmetry with time dependent sources\cite{MS2}
and for stationary problems in
absence of rotational symmetry \cite{MS3};
one could apply them to the problem of CTC
in 2+1 dimensions in
absence
of rotational symmetry and in presence of time dependence.

\end{document}